\documentclass[conference]{IEEEtran}

\usepackage{cite}
\usepackage[cmex10]{amsmath}
\usepackage{amsfonts,amssymb}
\usepackage[pdftex]{graphicx}
\usepackage[algoruled]{algorithm2e}
\usepackage{graphicx}
\usepackage[paper=letterpaper,margin=1in]{geometry} 

  \newtheorem{theorem}{Theorem}[section]
  \newtheorem{corollary}[theorem]{Corollary}
  \newtheorem{proposition}[theorem]{Proposition}
  \newtheorem{lemma}[theorem]{Lemma}

  \newtheorem{definition}[theorem]{Definition}
  
 \newtheorem{condition}{Condition}
 \newtheorem{remark}[theorem]{Remark}

\newcommand{\vc}[1]{{\mathbf{ #1}}}

\newcommand{\II}{{\mathbb{I}}}
\newcommand{\ep}{{\mathbb{E}}}
\newcommand{\pr}{{\mathbb{P}}}

\newcommand{\ol}[1]{\overline{#1}}

\newcommand{\suc}{{\rm suc}}
\newcommand{\tends}{\rightarrow \infty}
\newcommand{\setS}{{\mathcal{S}}}
\newcommand{\setP}{{\mathcal{P}}}
\newcommand{\setX}{{\mathcal{X}}}
\newcommand{\nec}{{\rm nec}}
\newcommand{\bd}{{\rm bd}}

\begin{document}

\title{The capacity of non-identical \\
adaptive group testing}
\date{\today}
\author{\IEEEauthorblockN{Tom Kealy} \IEEEauthorblockA{CDT in Communications   \\ MVB, School of Engineeering \\ University of Bristol, UK \\ Email: tom.kealy.kealy@bristol.ac.uk}
\and \and 
\IEEEauthorblockN{Oliver Johnson} \IEEEauthorblockA{School of Mathematics\\ University Walk, Bristol \\
University of Bristol, UK \\ Email: O.Johnson@bristol.ac.uk}
\and
\IEEEauthorblockN{ Robert Piechocki} \IEEEauthorblockA{CSN Group \\ MVB, School of Engineeering \\ University of Bristol, UK \\
Email: r.j.piechocki@bristol.ac.uk}
}
\maketitle

\begin{abstract}
\noindent We consider the group testing problem, in the case where the items are defective independently but with non-constant probability.
We introduce and analyse an algorithm to solve this problem by grouping items together appropriately. We give conditions under which  the
algorithm performs essentially optimally in the sense of information-theoretic capacity.
We use concentration of measure results to bound the probability that this algorithm requires many more tests than the expected number.
This has applications to the allocation of spectrum to cognitive radios, in the case where a database gives prior information that a particular band will be occupied.
\end{abstract}

\section{Introduction and notation}

\subsection{The Probabilistic group testing problem}
Group testing is a sparse inference problem, first introduced by Dorfman \cite{dorfman} in the context of testing for rare diseases.
Given  a large population of items $\setP$,
indexed by \( \{1, \ldots N\}\), where
some small fraction of the items are interesting in some way, how can we find the interesting items efficiently? 

We perform a sequence of $T$ pooled tests defined by test sets $\setX_1, \ldots, \setX_T$, where each $\setX_i \subseteq \setP$. 
We represent the interesting (`defective') items by 
a random vector $\vc{U} = ( U_1, \ldots, U_N)$, where $U_i$ is the indicator of the event that item $i$ is defective. 
For each test $i$, we jointly 
test all the items in $\setX_i$, and the outcome $y_i$ is `positive' ($y_i = 1$) if and only if any item in $\setX_i$ is defective. In other words, 
$y_i = \II \left( \sum_{j \in \setX_i} U_j \geq 1\right)$, since for simplicity we are  considering the noiseless case. Further, in this paper, we restrict our attention to
 the adaptive case, where we choose test set
$\setX_i$ based on a knowledge of sets $\setX_1, \ldots, \setX_{i-1}$ and outcomes $y_1, \ldots, y_{i-1}$. The group testing problem requires us to infer $\vc{U}$ with high
probability given a low number of tests $T$.

Since Dorfman's paper \cite{dorfman}, there has been considerable work on the question of how to design the sets $\setX_i$ in order to minimise the number of tests $T$
required. In this context, we briefly mention so-called combinatorial designs (see \cite{du, malyutov} for a summary, with \cite{malyutov} giving invaluable
references to an extensive body of Russian work in the 1970s and 1980s). Such designs typically aim to ensure that 
set-theoretic properties known as disjunctness
and separability occur. In contrast, for simplicity of analysis, as well as  performance of optimal order, it is possible to consider random designs. Here sets $\setX_i$ are chosen
at random, either using constructions such as independent Bernoulli designs \cite{atia, johnsonc8, johnson33} or more sophisticated  random designs based on LDPC codes \cite{wadayama}. 

Much previous work has focussed on the Combinatorial group testing problem, where there are a fixed number of defectives $K$, and the defectivity vector $\vc{U}$ is chosen uniformly among all
binary vectors of weight $K$. In contrast, in this paper we study a Probabilistic group testing problem as formulated
for example in the work of Li et al. \cite{li5}, in that we suppose 
 each item is defective  independently with probability \(p_i\), or equivalently take $U_i$ to be independent Bernoulli($p_i$).

This Probabilistic framework, including non-uniform priors, is natural for many applications of group testing.
For example, see \cite{atia2}, the cognitive radio problem can be formulated in terms of  a population
 of communication bands in frequency spectra with some (unknown) occupied bands you must not utilise. Here, the values of $p_i$ may be chosen
 based on some database of past spectrum measurements or other prior information. Similarly, as in Dorfman's original work  \cite{dorfman} or more recent 
research \cite{shental} involving screening for genetic conditions, values of
$p_i$ might summarise prior information based on a risk profile or  family history.
\subsection{Group testing capacity}
It is possible to characterize performance tradeoffs in group testing 
 from an information-theoretic point of view -- see for example \cite{atia, johnsonc10, johnson33, tan}. These papers have focussed on group testing as a channel coding problem, 
with \cite{atia, tan}
explicitly calculating the mutual information.
The paper \cite{johnsonc10} defined the capacity of a Combinatorial group testing procedure, which characterizes the number of bits of information
about the defective set which we can learn per test. We give a more general definition here,
which covers both the Combinatorial and Probabilistic cases.
\begin{definition} \label{def:capacity} Consider a sequence of group testing problems where the $i$th problem
has defectivity vector  $\vc{U}^{(i)}$, and consider algorithms which are given $T(i)$ tests.
We refer to a constant $C$ as the (weak) group testing capacity if for any $\epsilon > 0$:
  \begin{enumerate}
    \item any sequence of algorithms with
      \begin{equation} \label{eq:lower}
        \liminf_{i \tends} \frac{ H(\vc{U}^{(i)}) }{T(i)} \geq C+ \epsilon,
      \end{equation}
      has success probability $\pr(\suc)$ bounded away from 1,
    \item and there exists a sequence of algorithms with
      \begin{equation} \label{eq:upper}
        \liminf_{i \tends} \frac{H(\vc{U}^{(i)}) }{T(i)}  \geq C - \epsilon
      \end{equation}
      with success probability $\pr(\suc) \rightarrow 1$.
  \end{enumerate}
\end{definition}
\begin{remark}
In the Combinatorial case of $K$ defective items with all defective sets equally
likely, $ H(\vc{U}) = \log_2 \binom{N}{K}$, which is the term found in the denominator in \cite[Eq. (1) and (2)]{johnsonc10}. In the Probabilistic case (as in \cite{li5}) 
 we know $H(\vc{U}) = -\sum_{i=1}^N h(p_i)$ where $h(t) = -t \log_2 t - (1-t) \log_2(1-t)$ is the binary entropy function. 
\end{remark}
\begin{remark} If for $ \liminf_{i \tends} \frac{ H(\vc{U}^{(i)}) }{T(i)} \geq C+ \epsilon$, the success probability 
$\pr(\suc) \rightarrow 0$ we say that $C$ is the strong group testing capacity, following standard terminology in information
theory. Such a result is referred to as a strong converse.
\end{remark}

\subsection{Main results}

The principal contribution of \cite[Theorem 1.2]{johnsonc10} was the following result:
\begin{theorem}
[\cite{johnsonc10}] \label{thm:mainold}
The strong capacity of the adaptive noiseless Combinatorial group testing problem
  is  $C = 1$,
  in any regime such that $K/N \rightarrow 0$.
\end{theorem}
This argument came in two parts. First, in \cite[Theorem 3.1]{johnsonc10} the authors  proved a new
upper bound on success probability 
\begin{equation} \label{eq:bja} \pr(\suc) \leq \frac{2^T}{\binom{N}{K}}, \end{equation} which implied a strong converse ($C \leq 1$). This was complemented 
by 
showing that, in the Combinatorial case, an algorithm based on 
Hwang's Generalized Binary Splitting Algorithm (HGBSA) \cite{hwang}, \cite{du} is essentially optimal in the required sense,
showing that $C=1$ is achievable.

It may be useful to characterize the Probabilistic group testing problem in 
terms of the effective sparsity $\mu^{(N)}: = \sum_{i=1}^N p_i$. In particular, if the $p_i$ are (close to) identical, we would expect performance similar to that in the Combinatorial
case with $K = \mu^{(N)}$ defectives. As in \cite{johnsonc10}, we  focus on asymptotically sparse cases, where $\mu^{(N)}/N \rightarrow 0$. In contrast, Wadayama
\cite{wadayama} considered a model where $p_i$ are identical and fixed.
The main result of the present paper is Theorem \ref{thm:main}, stated and proved in Section \ref{sec:main} below, which implies the 
following  Probabilistic group testing version of Theorem \ref{thm:mainold}.
\begin{corollary} \label{cor:main}
 In the case where $p_i \equiv p$, the weak capacity of the adaptive noiseless Probabilistic group testing problem
  is  $C = 1$, in any regime such that $\mu^{(N)}/N \rightarrow 0$ and $\mu^{(N)} \rightarrow \infty$.
  \end{corollary}

Again we prove our main result Theorem \ref{thm:main} using complementary bounds on both sides. First in Section \ref{sec:ub} we recall
 a universal upper bound on success probability, Theorem \ref{thm:upper}, taken from
\cite{li5}, which implies a weak converse.   In \cite{li5}, Li et al. introduce
the Laminar Algorithm for Probabilistic group testing.  
In Section \ref{sec:algo} we propose a refined version of this Laminar Algorithm, based on Hwang's HGBSA \cite{hwang}, which is
analysed in Section \ref{sec:main}, and shown to imply performance close to optimal in the sense of capacity.

Bounds of \cite{li5} (see Theorem \ref{thm:lower} below) which shows that \(10eH\left(\textbf{U}\right)\) tests are required to guarantee convergence to zero success probability. Our calculation is to improve this to \(H\left(\textbf{U}\right)\) plus an error 
term, which is optimal up to the size of the error term.

\section{Algorithms and existing results}

\subsection{Upper bounds on success probability} \label{sec:ub}

Firstly \cite[Theorem 1]{li5} can be restated to give the following
upper bound on success probability:
\begin{theorem} \label{thm:upper}
Any Probabilistic group testing algorithm using $T$ tests with noiseless measurements has success probability satisfying
$$ \pr(\suc) \leq \frac{T}{H( \vc{U})}.$$
\end{theorem}

Rephrased in terms of Definition \ref{def:capacity}, this tells us that the weak capacity of noiseless Probabilistic 
group testing is $\leq 1$. The logic is as follows; if the capacity were
$1 + 2 \epsilon$ for some $\epsilon > 0$, then there would exist a sequence of algorithms with $H(\vc{U}^{(i)})/T(i) \geq 1 + \epsilon$ with success probability tending to 1.
However, by Theorem \ref{thm:upper}, any such algorithms have $\pr(\suc) \leq 1/(1+\epsilon)$, meaning that we have 
established that a weak converse holds.

\begin{remark}
It remains an open and interesting problem to prove an equivalent of \eqref{eq:bja} as in \cite[Theorem 3.1]{johnsonc10}. That is we hope to find an upper bound
on success probability in a form  which implies a strong converse, and hence that the strong capacity of Probabilistic group
testing is equal to 1.
\end{remark}

\subsection{Binary search algorithms}

The main contribution of this work is  to describe and analyse algorithms that will find the defective items.
In brief, we can think of Hwang's HGBSA algorithm as dividing the population $\setP$ into search sets $\setS$. First, all the items in a search set $\setS$ are tested together,
using a test set $\setX_1 = \setS$. If the result is negative
($y_1 = 0$), we can be certain that $\setS$ contains no defectives. However, if the result is positive ($y_1 = 1$),  $\setS$ must contain at least one defective.

If $y_i = 1$, we can be guaranteed to find at least
one defective, using the following binary search strategy.  
We  split the   set $\setS$ in two, and test the `left-hand' set, say $\setX_{2}$. If $y_{2} = 1$, then we know that $\setX_{2}$ contains at least one defective.
If $y_{2} = 0$, then $\setX_{2}$ contains no defective, so we can deduce that $\setS \setminus \setX_2$ contains at least one defective. By repeated use of this strategy, we are 
guaranteed to find a succession of nested sets which contain at least one defective, until $\setX_i$ is of size 1, and we have isolated a single defective item.

However this strategy may not find every defective item in $\setS$. To be specific, it is possible that at some stage
both the left-hand and right-hand sets contain a defective. The Laminar Algorithm of \cite{li5} essentially deals with this by testing 
both sets. However, we believe that this is inefficient, since typically both sets will not contain a defective. Nonetheless, the Laminar Algorithm satisfies
 the following  performance guarantees  proved in \cite[Theorem 2]{li5}: 
\begin{theorem} \label{thm:lower}
The expected number of tests required by the Laminar Algorithm \cite{li5}  is $\leq 2 H(\vc{U}) + 2 \mu$. Under a technical condition (referred to as non-skewedness), the
success probability can be bounded by $\pr(\suc) \geq 1- \epsilon$ using $T = (1+ \delta) (2^{\Gamma + \log_2 3} + 2) H(\vc{U})$ tests, where $\Gamma$ is defined implicitly
in terms of $\epsilon$, and $\delta \geq 2 e - 1$.
\end{theorem}

Ignoring the $\Gamma$ term, and assuming the non-skewedness condition holds,
 this implies that (using the methods of \cite{li5}) $T =  2 e (3 + 2) H(\vc{U}) = 10 e H(\vc{U})$ tests are required to guarantee convergence to $1$
of the success probability. In our language, this implies a lower bound of $C \geq 1/(10 e) = 0.0368$. Even ignoring the analysis of error probability, the fact that the expected number
of tests is $\leq 2 H(\vc{U}) + 2 \mu$ suggests that we cannot hope to achieve $C > 1/2$ using the Laminar Algorithm.
\subsection{Summary of our contribution} \label{sec:algo}

\begin{algorithm}
 \KwData{A Set \(S\) of \(\lvert S \rvert = n\) items, \(\mu\) of which are actually defective in expectation, a probability vector \( \vec{p}^{\left(n\right)} \) describing each item's independent probability of being defective, and a cutoff \(\theta\)}
 
 \KwResult{The set of defective items}
 
 Discard items with \(p_i \leq \theta\)
 \\
 Sort the remaining items into \(B\) bins, collecting items together with \(p_i \in \left[1/2\Gamma^r,1/2\Gamma^{r-1}\right)\) in bin \(r\). Bin \(0\) contains items with probability \( \geq 1/2\)
 \\
 Sort the items in each bin into sets s.t. the total probability of each set is \(\sim 1/2\).
 \\
 Test each set in turn
 \\
   	\If{The test is positive}{Arrange the items in the set on a Shannon-Fano/Huffman Tree and recursively search the set for all the defectives it contains}
\end{algorithm}

The main contribution of our paper is a refined version of the Laminar Algorithm, summarised above, and an analysis resulting in tighter error bounds as formulated in Proposition \ref{prop:overall} (in terms of expected number of tests) and Theorem \ref{thm:main} (in terms
of error probabilities).
The  key ideas are:
\begin{enumerate}
\item To partition the population $\setP$ into search  sets  $\setS$ containing items which have similar probabilities,
expressed through the Bounded Ratio Condition \ref{cond:ratio}. This is discussed in Section \ref{sec:boundedratio}, and optimised in the proof
of Proposition \ref{prop:overall}.
\item The way in which we deal with sets $\setS$ which contain more than one defective, as discussed in Remark \ref{rem:algo} below. Essentially we do not backtrack after each test by
testing both left- and right-hand sets, but only backtrack after each defective is found.
\item To discard items which have probability below a certain
threshold, since with high probability none of them will be defective. This is an idea introduced in \cite{li5} and discussed
in Section \ref{sec:discard}, with a new bound given in Lemma \ref{lem:thresh}.
\item  Careful analysis in Section \ref{sec:expectation} of the properties of search sets $\setS$ gives Proposition \ref{prop:overall}, which shows that the expected number of tests required can
be expressed as $H(\vc{U})$ plus an error term. In Section \ref{sec:main},
we give an analysis of the error probability using Bernstein's inequality, Theorem \ref{thm:bernstein}, allowing us to
prove Theorem \ref{thm:main}.
\end{enumerate}

\subsection{Wider context: sparse inference problems}

Recent work \cite{aksoylar,tan} has shown that many arguments and bounds hold in a common framework of sparse inference
which includes group testing and compressive sensing.

Digital communications, audio, images, and text are examples of data sources we can compress. We can do this, because these data sources are sparse: they have fewer degrees of freedom than the space they are defined upon. 
For example, images have a well known expansion in either the Fourier or Wavelet bases. The text of an English document will only be comprised of words from the English dictionary, and not all the possible strings from the space of strings made up from the characters \(\{a, \ldots, z \}\). 

Often, once a signal has been acquired it will be compressed. However, the compressive sensing paradigm introduced by
\cite{candes,donoho2} shows that this isn't necessary. In those papers it was shown that a 'compressed' representation of a signal could be obtained from random linear projections of the signal and some other basis (for example White Gaussian Noise). The question remains, given this representation how do we recover the original signal? For real signals, a simple linear programme suffices.
Much of the work in this area has been couched in terms of the sparsity of the signal and the various bases the signal can be represented in (see for example \cite{candes,donoho2}).

\section{Analysis and new bounds}

\subsection{Searching a set of bounded ratio} \label{sec:boundedratio}

Recall that we have a population $\setP$ of items to test, each with associated probability of defectiveness $p_i$.
The strategy of the proof is to partition $\setP$ 
 into search sets $\setS_1, \ldots, \setS_G$, each of which contains items which have
comparable values of $p_i$.

\begin{condition}[Bounded Ratio Condition] \label{cond:ratio}
Given $\Gamma \geq 1$, say that a set $\setS$ satisfies the Bounded Ratio Condition with constant $\Gamma$ if
\begin{equation} \label{eq:ratio} \max_{i,j \in \setS} \frac{p_j}{p_i} \leq \Gamma.\end{equation} \end{condition}
(For example clearly if $p_i \equiv p$, any set $\setS$ satisfies the condition for any $\Gamma \geq 1$).
\begin{lemma} \label{lem:sfstep}
Consider a set $\setS$ satisfying the Bounded Ratio Condition  with constant $\Gamma$
and write $P_{\setS} = \sum_{j \in \setS} p_j$.
In a  Shannon--Fano tree for the probability distribution $\ol{p}_i := p_i/P_{\setS}$, each item has length $\ell_i^{(\setS)}$
bounded by
\begin{equation} \label{eq:depth} \ell_i^{(\setS)} \leq \ell_{\max}^{(\setS)} := \frac{h(\setS)}{P_{\setS}} + \log_2  \Gamma + \log_2  P_{\setS} + 1,\end{equation} 
where we write $h( \setS) :=  -\sum_{j \in \setS} p_j \log_2  p_j$.
\end{lemma}
\begin{IEEEproof} Under the Bounded Ratio Condition, for any $i$ and $j$, we know that by taking logs of \eqref{eq:ratio}
$$ -\log_2 p_i \leq - \log_2 p_j + \log_2 \Gamma.$$
Multiplying by $p_j$ and summing over all $j \in \setS$, we obtain that
\begin{equation} \label{eq:setbd} -P_{\setS} \log_2 p_i \leq h(\setS) + P_{\setS} \log_2 \Gamma.\end{equation}
Now, the Shannon--Fano length of the $i$th item is

\begin{eqnarray}
 \ell_i^{(\setS)} &= \lceil -\log_2 \ol{p}_i \rceil \\
&\leq &  -\log_2 p_i + \log_2 P_{\setS} + 1 \label{eq:lengthbd} \\
& \leq  & \left(\frac{h(\setS)}{P_{\setS}} + \log_2 \Gamma \right) + \log_2 P_{\setS} + 1. \nonumber
\end{eqnarray}

and the result follows by \eqref{eq:setbd}.
\end{IEEEproof}

Next we describe our search strategy:

\begin{remark} \label{rem:algo}
Our version of the algorithm will find every defective in a set $\setS$. We start as before by testing every item in $\setS$ together. If this test is negative, we are done.
Otherwise, if it is positive, we can perform binary search as section II-B to find one defective item, say $d_1$. Now, test every item in $\setS \setminus \{ d_1 \}$ together.
If this test is negative, we are done, otherwise we repeat the search step on this smaller set, to find another defective item
$d_2$, then we test $\setS \setminus \{ d_1, d_2 \}$ and so on.

We think of the algorithm as repeatedly searching a binary tree. Clearly, if the tree has depth bounded by $\ell$, then the search will take $ \leq \ell$ tests to find one defective. In total, 
if the set contains $U$ defectives, we need to repeat $U$ rounds of searching, plus the final test to guarantee that the set contains no more defectives, so will use $\leq \ell U + 1$ tests.
\end{remark}

\begin{lemma} \label{lem:expset}
 Consider a search set $\setS$ satisfying the Bounded Ratio Condition  and write $P_{\setS} = \sum_{j \in \setS} p_j$.
If (independently) item $i$ is defective with probability $p_i$, 
we can recover  all defective items in the set using $T_\setS$ tests, where $\ep T_\setS \leq T_{\bd}(\setS)$ for
\begin{equation} \label{eq:tbds}
 T_{\bd}(\setS) :=  h(\setS) +  P_{\setS} \log_2 \Gamma +  P_{\setS} \log_2 P_{\setS} + P_\setS + 1. \end{equation}
\end{lemma}
\begin{IEEEproof}
Using the algorithm of Remark \ref{rem:algo}, laid out on the Shannon-Fano tree constructed in Lemma \ref{lem:sfstep}, we are guaranteed  to find every defective.
The number of tests to find one defective thus corresponds to the depth of the tree, which is bounded by  $\ell_{\max}^{(\setS)}$ given in \eqref{eq:depth}.

Recall that we write $U_i$ for the indicator of the event that the $i$th item is defective, we will write $U_{\setS} = \sum_{i \in \setS} U_i$ for the total number of defectives in \(\setS\), and 
$l_i^{(\setS)}$ for the length of the word in the Shannon Fano tree. As discussed in Remark \ref{rem:algo}
this search procedure will take 
\begin{eqnarray}
T_\setS & = &  1 + \sum_{i \in \setS} U_i \ell_i^{(\setS)} \nonumber \\
& = &  \sum_{i \in \setS} p_i \ell_i^{(\setS)} + 1 + \sum_{i \in S} \ell_i^{(\setS)} (U_i - p_i)  \nonumber \\
& \leq & \sum_{i \in \setS} p_i \ell_{\max}^{(\setS)} +  1 + \sum_{i \in \setS} V_i^{(\setS)}  \nonumber \\
& =  &  P_{\setS} \ell_{\max}^{(\setS)} +  1 + \sum_{i \in S} V_i^{(\setS)} \nonumber \\
& \leq  & T_{\bd}(\setS) + \sum_{i \in S} V_i^{(\setS)} 
\mbox{ \;\; tests.} \label{eq:testsperdef} \end{eqnarray}
Here we write $V_ i^{(\setS)} = \ell_i^{(\setS)} (U_i - p_i)$, which has expectation zero, and \eqref{eq:testsperdef} follows using the expression for 
$\ell_{\max}^{(\setS)}$ given in   Lemma \ref{lem:sfstep}.
\end{IEEEproof}

\subsection{Discarding low probability items} \label{sec:discard}

As in \cite{li5}, we use a probability threshold $\theta$, and write $\setP^*$ for the population having removed items with $p_i \leq \theta$.
If an item lies in $\setP \setminus \setP^*$ we do not 
test it, and simply mark it as non-defective. This truncation operation gives an error if and only if some item in $\setP \setminus \setP^*$ is defective.
By the union bound, 
this truncation operation contributes a total of $\pr( \mbox{$\setP \setminus \setP^*$ contains a defective}) \leq  \rho := \sum_{i=1}^n p_i \II(p_i \leq \theta)$ to the error probability.

\begin{lemma} \label{lem:thresh}
Choosing $\theta(P_e)$ such that  
\begin{equation} \label{eq:thetadef}
-\log_2 \theta(P_e) = \min\left( \log_2  \left( \frac{2n}{P_e} \right), \frac{2 H( \vc{U})}{P_e} \right)
\end{equation}
ensures that  
\begin{equation}
\pr(\mbox{ $\setP \setminus \setP^*$ contains a defective} ) \leq P_e/2. \label{eq:setpstar} \end{equation} \end{lemma}
\begin{IEEEproof}
The approach of \cite{li5} is essentially to bound $\II(p_i \leq \theta) \leq \theta/p_i$ so that
$\rho = \sum_{i=1}^n p_i \II(p_i \leq \theta) \leq \sum_{i=1}^n p_i (\theta/p_i) = n \theta$. Hence, choosing a threshold
of $\theta = P_e/(2n)$ guarantees the required bound on $\rho$.

We combine this with another bound, constructed using a different function:
 $\II(p_i \leq \theta) $ $\leq (-\log_2 p_i)/(-\log_2 \theta)$, so that
$$ \rho = \sum_{i=1}^n p_i \II(p_i \leq \theta) \leq \sum_{i=1}^n p_i  \left( 
\frac{-\log_2 p_i}{-\log_2 \theta } \right) \leq \frac{H( \vc{U})}{-\log_2 \theta},$$ 
so we deduce the result. \end{IEEEproof}

\subsection{Searching the entire set} 

Having discarded items with $p_i$ below
this probability threshold $\theta$ and given bounding ratio $\Gamma$,
we create a series of bins. We collect together items with probabilities $p \in [1/2,1]$ in bin 0,
$p \in [1/(2\Gamma), 1/2)$ in bin 1, items with probabilities $p \in [1/(2\Gamma^2), 1/(2\Gamma))$ in bin 2, \ldots, and items with probabilities $p \in 
[1/(2\Gamma^B), 1/(2\Gamma^{B-1}))$ in bin $B$.

The probability threshold $\theta$ means that
there will be a finite number of such bins, with the index $B$ of the last bin defined by the fact that $1/(2\Gamma^B) \leq
\theta < 1/(2\Gamma^{B-1})$, meaning that $(B -1) \log_2 \Gamma < - \log_2 (2 \theta)$, so
\begin{equation} \label{eq:bincount} B \leq  \frac{ - \log_2 (2 \theta)}{\log_2 \Gamma} + 1. \end{equation}

We split the items in each bin into search sets $\setS_i$, motivated by the following definition:

\begin{definition} \label{def:full}
A set of items \( \setS \) is said to be \(full\) if $P_{\setS} = \sum_{i \in \setS} p_i \geq \frac{1}{2}$.
\end{definition}

Our splitting procedure is as follows: we create a list of
possible sets $\setS_1, \setS_2, \ldots $. For $i$ increasing from $0$ to $B$,
we place items from bin $i$ into sets $\setS_{b_{i}+1}, \ldots, \setS_{b_{i+1}}$, for some $b_i$, where $b_{0} = 0$.
Taking the items 
from bin $i$, while $\setS_{b_{i}+1}$ is not full  (has 
total probability $< \frac{1}{2}$) we will place  items into it. Once enough items have been added to fill $\setS_{b_{i} + 1}$,
we will proceed in the same way to fill $\setS_{b_{i}+2}$, and so on until all the items in bin $i$ have been
assigned to sets $\setS_{b_{i}+1}, \ldots, \setS_{b_{i+1}}$,  where $\setS_{b_{i+1}}$ may remain not full.
\begin{proposition} \label{prop:splitting}
 This splitting procedure will divide $\setP^*$ into search sets $\setS_1, \ldots, \setS_G$, where
the total number of sets is 
$$ G \leq 2 \mu + B \leq 2 \mu  +  \left( \frac{ -\log_2 (2\theta)}{\log_2 \Gamma} + 1 \right).$$ 
Each set $\setS_j$ satisfies the Bounded Ratio Condition and has total probability $P_j := P_{\setS_j} \leq 1$. \end{proposition}
\begin{IEEEproof}
First, note that  
the items from bin $0$ each lie in a set $\setS$ on their own.
These sets will be full, trivially satisfy the Bounded Ratio Condition \ref{cond:ratio}
with constant $\Gamma$,
 and have probability satisfying $P_j \leq 1$.
 For each of bins $1, \ldots, B$:
\begin{enumerate}
\item  \label{it:count} For each bin $i$,
it is possible that the last set $\setS_{b_{i+1}}$  will not be full, but every other
set corresponding to that bin will be full. Hence, there are no more than $B$ sets which are not full.
\item For each resulting set $\setS_j$, the total probability $P_j \leq 1$ (since just before we add the final item, $\setS_j$ is not full, so at
that stage has total probability $\leq 1/2$, and each element in bins $1, \ldots, B$ has probability $\leq 1/2$).
\item Since each set $\setS_j$ contains items  taken from the same bin, it will satisfy the Bounded Ratio Condition  with 
constant $\Gamma$.
\end{enumerate}

 Note that the number of full sets is \(\leq 2 \mu \), since
\begin{align} 
\label{eq:counting}
\mu &=  \sum_{i \in \setP} p_i \\ 
&\geq \sum_{i \in \setP^*} p_i  = \sum_{j=1}^G P_j
\\ &\geq
\sum_ {\mbox{\scriptsize $j$: $\setS_j$ full}} P_j \geq \left| \mbox{$\setS_j$ full} \right| \frac{1}{2}. \end{align}
Since, as discussed  in point \ref{it:count}) above,
 the total number of sets is bounded by the number of full sets plus $B$, the result follows using Equation
(\ref{eq:bincount}).
\end{IEEEproof}

\subsection{Bounding the expected number of tests} \label{sec:expectation}
We allow the algorithm to work until all defectives in $\setP^*$ are found, and write $T$ for the (random) number of tests this takes.
\begin{proposition} \label{prop:overall} Given a population $\setP$
where (independently) item $i$ is defective with probability $p_i$,  
we  recover  all defective items in $\setP^*$ in $T$ tests with $\ep T \leq T_{\bd}$, where
\begin{equation} \label{eq:tbd}
 T_{\bd} := \left( H( \vc{U}) + 3 \mu + 1 \right) +  
 2 \sqrt{ \mu  \left( -\log_2 (2\theta) \right)}.
\end{equation}
\end{proposition}
\begin{IEEEproof} Given a value of $\Gamma$,
Proposition \ref{prop:splitting} shows that our splitting procedure divides $\setP^*$ into 
$G$ sets $\setS_1, \ldots, \setS_G$, such that each set $\setS_j$ satisfies the Bounded Ratio Condition with constant $\Gamma$ and has total probability $P_j
\leq 1$. Using the notation of Lemma \ref{lem:expset}, $T = \sum_{j=1}^G T_{\setS_j}$, where 
$\ep T_{\setS_j} \leq  T_{\bd}(\setS_j)$.

Adding this bound over the different sets, since $P_j \leq 1$ means that $P_j \log_2 P_j \leq 0$, we obtain
\begin{align} \label{eq:total}
\lefteqn{ \sum_{j=1}^G T_{\bd}(\setS_j) } \nonumber \\
&\leq \sum_{j=1}^G \left(  h(\setS_j)  + P_j (\log_2 \Gamma+1) + 1 \right) \nonumber \\
&= \sum_{j \in \setP^*} -p_j \log_2 p_j + \mu  (\log_2 \Gamma+1) + G \nonumber \\
&\leq \sum_{j \in \setP^*} h(p_j)  + 3 \mu  +  1 + \left( \frac{ -\log_2 (2\theta)}{\log_2 \Gamma} +  \mu  \log_2 \Gamma \right)  
\nonumber \\
&\leq \left( H( \vc{U}) + 3 \mu + 1 \right) \\ 
&+  \left( \frac{ -\log_2 (2 \theta)}{\log_2 \Gamma} +  \mu \log_2 \Gamma \right).
\label{eq:toopt}
\end{align}
This follows by the bound on $G$ in Proposition \ref{prop:splitting}, as well as 
the fact that $ 0 \leq p_j \leq 1$ means that 
 for any $i$, $ - p_j \log_2 p_j =  (1-p_j) \log_2 (1-p_j) + h(p_j) \leq h(p_j)$.

Finally, we  choose $\Gamma > 1$ to optimize the  second bracketed term in Equation (\ref{eq:toopt}).
Differentiation shows that the optimal $\Gamma$ satisfies $\log_2 \Gamma = \sqrt{ -\log_2 (2 \theta)/\mu},$ meaning that
 the 
bracketed term $$
\frac{ -\log_2 (2 \theta)}{\log_2 \Gamma} +  \mu \log_2 \Gamma  =
 2 \sqrt{ \mu  \left( -\log_2 (2\theta) \right)} ,$$
and the result follows.
\end{IEEEproof}

\subsection{Controlling the error probabilities} \label{sec:main}
Although Section \ref{sec:expectation} proves that $\ep T \leq T_{\bd}$,
 to bound the capacity, we need to prove that with high probability
$T$ is not significantly larger than $T_{\bd}$. This can be done using
 Bernstein's inequality (see for example Theorem 2.8 of \cite{petrov}):
\begin{theorem}[Bernstein] \label{thm:bernstein}
For zero-mean random variables $V_i$ which are uniformly bounded by $|V_i| \leq M$,
if we write $L  := \sum_{j=1}^n \ep V_j^2 $
 then
\begin{equation} \pr \left( \sum_{j=1}^n V_j \geq t \right)  \leq  \exp \left( - \frac{ t^2}{4 L}  \right),  \mbox{
for any $0 \leq t \leq \frac{L}{M}$.}   \label{eq:bernstein}
\end{equation}
\end{theorem}
We deduce the following result:
\begin{theorem} \label{thm:main}
Write  $L = \sum_{j \in \setP^*} l_j^2 p_j (1-p_j)$,  $M = -\log_2 \theta+1$
and $\psi = (L/(4M^2))^{-1/3}$.
Define
\begin{equation} \label{eq:tnec}
 T_{\nec} = T_{\bd} + \psi H(\vc{U}), \end{equation}
where $T_{\bd} $ is given in \eqref{eq:tbd}.
\begin{enumerate}
\item \label{it:part1} If we terminate our group testing algorithm after
$T_{\nec}$ tests, the success probability
\begin{equation} \label{eq:errorprob} 
\pr(\suc) \geq 1 - \frac{1}{2} \sqrt{ \frac{\mu}{H(\vc{U}) }} -  \exp \left( - \left( \frac{L}{4  M^2 }  \right)^{1/3}
\right).
\end{equation}
\item \label{it:part2}
Hence  in any regime where $\mu \rightarrow \infty$ with
 $\mu/H(\vc{U}) \rightarrow 0$ and $L/M^2 \rightarrow \infty$,  
our group testing algorithm has  (a) $ \liminf H(\vc{U})/T_{\nec} \geq 1/(1+ \epsilon)$ for any $\epsilon$ and (b)
$\pr(\suc) \rightarrow 1$, so the capacity $C = 1$.
\end{enumerate}
\end{theorem}
\begin{IEEEproof}
We first prove the success probability bound \eqref{eq:errorprob}.
 Recall that our algorithm searches the reduced population set $\setP^*$ for defectives.
This gives two  error events -- either there are defective items in the set $\setP \setminus \setP^*$, or the algorithm does not find all the defectives in $\setP^*$ using
$T_{\nec}$ tests. We consider them separately, and control the probability of either happening using the union bound.

Writing $H = H(\vc{U})$  for brevity and
choosing $ P_e = \sqrt{ \mu/H }$ ensures that (by Lemma \ref{lem:thresh}) the first event has probability $\leq P_e/2$, contributing 
$\frac{1}{2} \sqrt{ \mu/H(\vc{U})}$ to (\ref{eq:errorprob}).

Our analysis of the second error event is based on  the 
random term   from
Equation \eqref{eq:testsperdef}, which we previously averaged over but now wish to bound. There will be an error
if $T_{\nec} \leq T$, or (rearranging) if 
$$\psi H \leq T - T_{\bd} \leq \sum_{j=1}^G \left(T_{\setS_j} - T_{\bd}(\setS_j) \right) = \sum_{i \in \setP^*} V_i.$$  For brevity,   for $i \in \setS$, we write 
$V_i = V_ i^{(\setS)} = \ell_i^{(\setS)} (U_i - p_i)$ and $\ell_i = \ell_i^{(\setS)}$, where $V_i$ has expectation zero. 

We have discarded elements with probability below $\theta$, as given by \eqref{eq:thetadef}, and by design all the sets $\setS$ have total probability $P_{\setS} \leq 1$. Using
 \eqref{eq:lengthbd} we know that the $V_i$ are bounded by
\begin{equation} | V_i|  \leq \ell_i \leq -\log_2 p_i + \log_2 P_{\setS} + 1 \leq - \log_2 \theta + 1. \label{eq:vbd} \end{equation}
Hence, the conditions of Bernstein's inequality, Theorem \ref{thm:bernstein}, are satisfied. 
Observe that since all $l_j \leq M$,
$$ \frac{ L}{H M} = \frac{ \sum_{j \in \setP^*} l_j^2 p_j (1-p_j)} { H M} \leq \frac{ \sum_{j \in \setP^*} l_j p_j } { H} \leq 1 .$$
Hence Theorem \ref{thm:bernstein} gives that 
\begin{eqnarray*}
 \pr \left( \sum_{j \in \setP^*} V_j \geq \psi H \right) 
& \leq &
 \pr \left( \sum_{j \in \setP^*} V_j \geq \psi L/M \right) \\
& \leq &  \exp \left( - \frac{L \psi^2}{4  M^2 } \right) \\ 
& = & \exp \left( - \left( \frac{L}{4  M^2 }  \right)^{1/3}
\right).
\end{eqnarray*}
Using the union bound, the probability bound \eqref{eq:errorprob} follows.

We next consider the capacity bound of \ref{it:part2}).
Since $-\log_2 \theta \leq 2 H/P_e$, using \eqref{eq:tbd}  and \eqref{eq:tnec}
\begin{eqnarray}
\frac{T_{\nec}}{H} & = & 
 \frac{T_{\bd}}{H}  + \psi \nonumber\\
&  = & 1 + 3 \frac{\mu}{H} + \frac{1}{H}  + 2 \sqrt{ \frac{ \mu}{H P_e}} + \psi \nonumber \\
& =  & 1 + 3 \frac{\mu}{H} + \frac{1}{H} + 2 \left( \frac{\mu}{H} \right)^{1/4} + \psi, \label{eq:ratio2}
\end{eqnarray}
which in our regime of interest is $\leq 1 + \epsilon$ in the limit, since \(\Phi = \left(L/ \left(4M^2\right)\right) \rightarrow 0\) by assumption.
\end{IEEEproof}

\begin{IEEEproof}[Proof of Corollary \ref{cor:main}]
In the case where all $p$ are identical, $\mu = N p$, $H  = N p (-\log p)$, so $\mu/H = 1/(-\log p) \rightarrow 0$. Similarly, $L = N p (-\log_2 p)^2$ and $M = ( -\log_2 p)$
so that $L/M^2 = N p \rightarrow \infty$ as required.
\end{IEEEproof}

\section{Results}
The performance of Algorithm 1 (in terms of the sample complexity) was analysed by simulating 500 items, with a mean number of defectives equal to 8, i.e. \(N = 500\) and \(\mu^{(N)} = 8\). 

The probability distribution \(\vc{p}\) was generated by a Dirichlet distribution with parameter $\alpha$.
This produces an output distribution whose uniformity can be controlled via the parameter \(\alpha\), as opposed to simply choosing a set of random numbers and normalise by the sum. Consider the case of two random numbers, \(\left(x,y\right)\), distributed uniformly on the square \(\left[0,1\right]^2\). Normalising by the sum \(\left(x+y\right)\) projects the point \(\left(x,y\right)\) onto the line \(x+y=1\) and so favours points closer to \((0.5,0.5)\) than the endpoints. The Dirichlet distribution avoids this by generating points directly on the simplex.

\begin{figure}[h]
\centering
\includegraphics[width=0.5\textwidth]{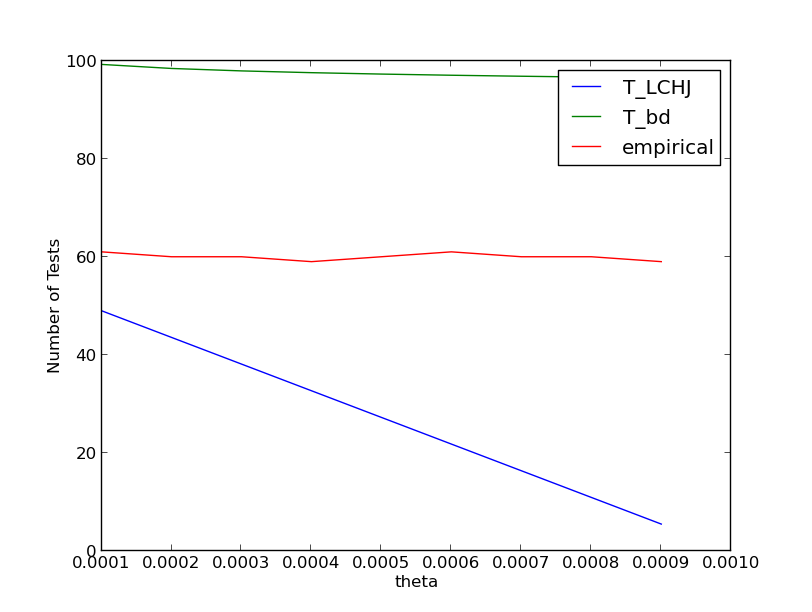}
\caption{Theoretical lower and upper bounds and empirical Test frequencies as functions of \(\theta\)}
\label{ubvslb}
\end{figure}

We then chose values of the cutoff parameter \(\theta\) from 0.0001 to 0.01, and for each \(\theta\) simulated the algorithm 1000 times. We plot the empirical distribution of tests, varying \(\theta\) as well as the uniformity/concentration of the probability distribution (via the parameter \(\alpha\) of the Dirichlet distribution). We also plot (in figure (\ref{ubvslb}), the theoretical lower and upper bounds on the number of Tests required for successful recovery alongside the empirical number tests (all as a function of \(\theta\)).

Note that the Upper bound is not optimal and there still is some room for improvement. Note also that the lower bound degrades with \(\theta_i \). The lower bound (\(T_{LCHJ}\)) was generated according to Theorem (\ref{thm:upper}). 

\begin{figure}[h]
\centering
\includegraphics[width=0.5\textwidth]{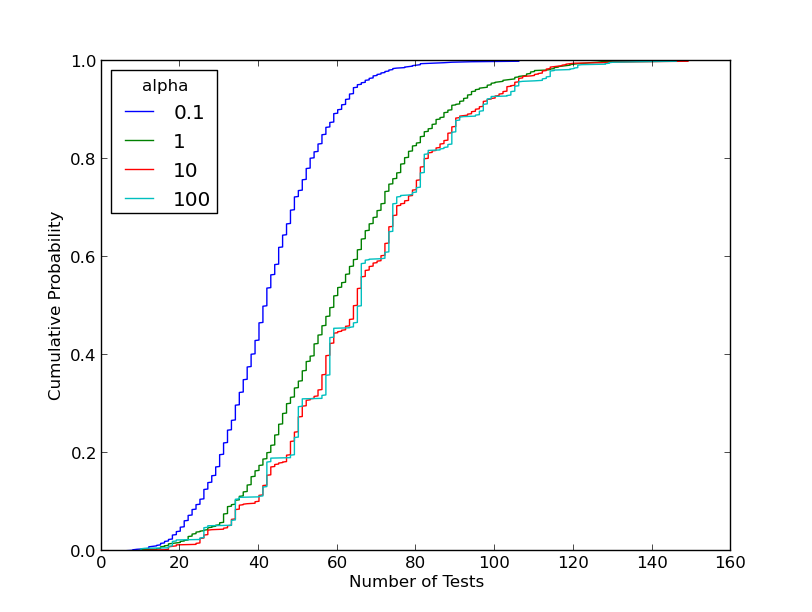}
\caption{Cumulative success probability distribution curves of the algorithm with fixed \(\theta = 0.0001\) and varying \(\alpha\) }
\label{testsvsalpha}
\end{figure}

\begin{figure}[h]
\centering
\includegraphics[width=0.5\textwidth]{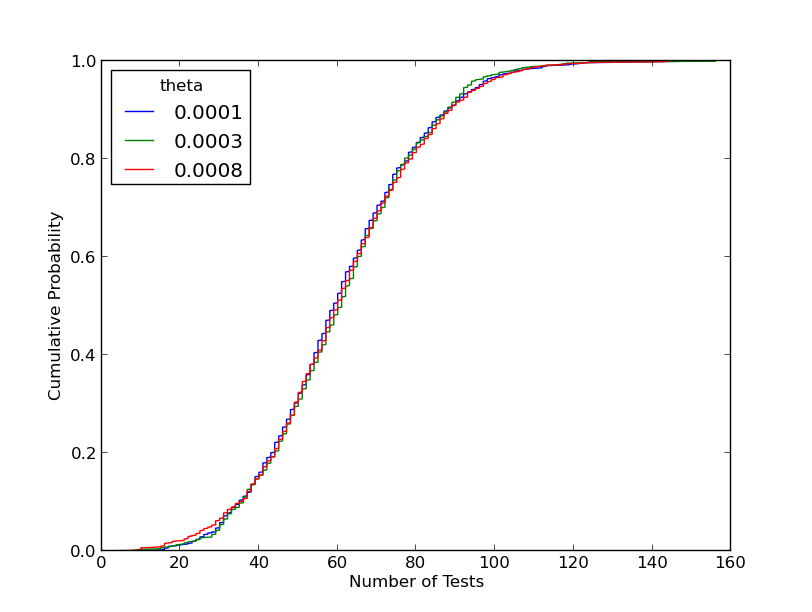}
\caption{Cumulative success probability distribution curves for fixed \(\alpha = 1\) and varying \(\theta\)}
\label{testsvstheta}
\end{figure}

Figures (\ref{testsvsalpha}) and (\ref{testsvstheta}) show that the performance is relatively insensitive to the cut-off \(\theta\), and more sensitive to the uniformity (or otherwise) of the probability distribution \(\vc{p}\). Heuristically, this is for because distributions which are highly concentrated on a few items algorithms can make substantial savings on the testing budget by testing those highly likely items first (which is captured in the bin structure of the above algorithm). 

The insensitivity to the cutoff \(\theta\) is due to items below \(\theta\) being overwhelmingly unlikely to be defective - which for small \(\theta\) means that few items (relative to the size of the problem) get discarded.

\section{Discussion}
We have introduced and analysed an algorithm for Probabilistic group testing which uses `just over' $H(\vc{U})$ tests to
recover all the defectives with high probability. Combined with a weak converse taken from \cite{li5}, this allows us to deduce that
the weak capacity of Probabilistic group testing is $C=1$.  
These results are illustrated by simulation.

For simplicity, this work has concentrated on establishing a bound $T_{\bd}$ in \eqref{eq:tbd} which has leading term $H(\vc{U})$,
and not on tightening bounds on the coefficient of $\mu$ in \eqref{eq:tbd}. For completeness, we mention that this coefficient
can be reduced from 3, under a simple further condition:

\begin{remark}
For some $c \leq 1/2$, we assume that all the $p_i \leq c$, and we alter the definition of `fullness' to assume that a set is
full if it has total probability less than $\alpha$. In this case, the term $P_{\setS} \log_2 P_{\setS}$ in \eqref{eq:tbds}
becomes $P_{\setS} \log_2 (\alpha + c)$, the bound in \eqref{eq:counting} becomes $\mu/\alpha$, and since
$\left( (1-p) \log_2 (1-p) \right)/p$ is decreasing in $p$, we can add a term $(1-c) \log_2 (1-c) $ to \eqref{eq:toopt}.
Overall, the coefficient of $\mu$ becomes $f(a,c) := \log_2 (\alpha + c) + 1 + 1/\alpha + (1-c) \log_2(1-c)$, which we can optimize over
$\alpha$. For example, if $c = 1/4$, taking $\alpha = 0.88824$, we obtain $f(a,c) = 2.00135$.
\end{remark}

It remains of interest to tighten the upper bound of Theorem \ref{thm:upper},
in order prove a strong converse, and hence confirm that the strong capacity is also equal to $1$.

In future work, we hope to explore more realistic models of defectivity, such as those where the defectivity of $U_i$ are not
necessarily independent, for example by imposing a Markov neighbourhood structure.

\section*{Acknowledgments}

This work was supported by the Engineering and Physical Sciences Research Council [grant number {\tt EP/I028153/1}]; Ofcom; and the University of Bristol. The authors would particularly like to thank Gary Clemo of Ofcom for useful discussions.

\bibliography{../../}

\begin{thebibliography}{10}
\providecommand{\url}[1]{#1}
\csname url@samestyle\endcsname
\providecommand{\newblock}{\relax}
\providecommand{\bibinfo}[2]{#2}
\providecommand{\BIBentrySTDinterwordspacing}{\spaceskip=0pt\relax}
\providecommand{\BIBentryALTinterwordstretchfactor}{4}
\providecommand{\BIBentryALTinterwordspacing}{\spaceskip=\fontdimen2\font plus
\BIBentryALTinterwordstretchfactor\fontdimen3\font minus
  \fontdimen4\font\relax}
\providecommand{\BIBforeignlanguage}[2]{{%
\expandafter\ifx\csname l@#1\endcsname\relax
\typeout{** WARNING: IEEEtran.bst: No hyphenation pattern has been}%
\typeout{** loaded for the language `#1'. Using the pattern for}%
\typeout{** the default language instead.}%
\else
\language=\csname l@#1\endcsname
\fi
#2}}
\providecommand{\BIBdecl}{\relax}
\BIBdecl

\bibitem{dorfman}
R.~Dorfman, ``The detection of defective members of large populations,''
  \emph{The Annals of Mathematical Statistics}, pp. 436--440, 1943.

\bibitem{du}
D.~Du and F.~Hwang, \emph{Combinatorial Group Testing and Its Applications},
  ser. Series on Applied Mathematics.\hskip 1em plus 0.5em minus 0.4em\relax
  World Scientific, 1993.

\bibitem{malyutov}
M.~Malyutov, ``Search for sparse active inputs: a review,'' in
  \emph{Information Theory, Combinatorics and Search Theory}, ser. Lecture
  notes in Computer Science.\hskip 1em plus 0.5em minus 0.4em\relax London:
  Springer, 2013, vol. 7777, pp. 609--647.

\bibitem{atia}
G.~Atia and V.~Saligrama, ``Boolean compressed sensing and noisy group
  testing,'' \emph{IEEE Trans. Inform. Theory}, vol.~58, no.~3, pp. 1880
  --1901, March 2012.

\bibitem{johnsonc8}
D.~Sejdinovic and O.~T. Johnson, ``Note on noisy group testing: Asymptotic
  bounds and belief propagation reconstruction,'' in \emph{Proceedings of the
  48th Annual Allerton Conference on Communication, Control and Computing},
  2010, pp. 998--1003.

\bibitem{johnson33}
M.~P. Aldridge, L.~Baldassini, and O.~T. Johnson, ``Group testing algorithms:
  bounds and simulations,'' \emph{IEEE Trans. Inform. Theory}, vol.~60, no.~6,
  pp. 3671--3687, 2014.

\bibitem{wadayama}
T.~Wadayama, ``An analysis on non-adaptive group testing based on sparse
  pooling graphs,'' in \emph{2013 IEEE International Symposium on Information
  Theory}, 2013, pp. 2681--2685.

\bibitem{li5}
T.~Li, C.~L. Chan, W.~Huang, T.~Kaced, and S.~Jaggi, ``Group testing with prior
  statistics,'' 2014, see {\tt arxiv:1401.3667}.

\bibitem{atia2}
G.~Atia, S.~Aeron, E.~Ermis, and V.~Saligrama, ``On throughput maximization and
  interference avoidance in cognitive radios,'' in \emph{Consumer
  Communications and Networking Conference, 2008. CCNC 2008. 5th IEEE}.\hskip
  1em plus 0.5em minus 0.4em\relax IEEE, 2008, pp. 963--967.

\bibitem{shental}
N.~Shental, A.~Amir, and O.~Zuk, ``Identification of rare alleles and their
  carriers using compressed se (que) nsing,'' \emph{Nucleic acids research},
  vol.~38, no.~19, pp. e179--e179, 2010.

\bibitem{johnsonc10}
L.~Baldassini, O.~T. Johnson, and M.~P. Aldridge, ``The capacity of adaptive
  group testing,'' in \emph{2013 IEEE International Symposium on Information
  Theory, Istanbul Turkey, July 2013}, 2013, pp. 2676--2680.

\bibitem{tan}
V.~Tan and G.~Atia, ``Strong impossibility results for sparse signal
  processing,'' \emph{IEEE Signal Processing Letters}, vol.~21, no.~3, pp.
  260--264, March 2014.

\bibitem{hwang}
F.~K. Hwang, ``A method for detecting all defective members in a population by
  group testing,'' \emph{Journal of the American Statistical Association},
  vol.~67, no. 339, pp. 605--608, 1972.

\bibitem{aksoylar}
C.~Aksoylar, G.~Atia, and V.~Saligrama, ``Sparse signal processing with linear
  and non-linear observations: A unified {S}hannon theoretic approach,'' in
  \emph{2013 IEEE Information Theory Workshop (ITW)}, Sept 2013, pp. 1--5.

\bibitem{candes}
E.~J. Cand{\`e}s, J.~Romberg, and T.~Tao, ``Robust uncertainty principles:
  Exact signal reconstruction from highly incomplete frequency information,''
  \emph{IEEE Trans. Inform. Theory}, vol.~52, no.~2, pp. 489--509, 2006.

\bibitem{donoho2}
D.~L. Donoho, ``Compressed sensing,'' \emph{IEEE Trans. Inform. Theory},
  vol.~52, no.~4, pp. 1289--1306, 2006.

\bibitem{petrov}
V.~V. Petrov, \emph{Limit Theorems of Probability Theory: Sequences of
  Independent Random Variables}.\hskip 1em plus 0.5em minus 0.4em\relax Oxford:
  The Clarendon Press, 1995.

\end{thebibliography}
\end{document}